\documentclass[prl,showpacs,twocolumn,aps,superscriptaddress,a4paper]{revtex4-1}

\usepackage{pst-node}
\usepackage{amsmath}
\usepackage{graphicx}
\usepackage{latexsym}
\usepackage{amsfonts}
\usepackage{amssymb}
\usepackage{graphicx}
\usepackage{rotating}
\usepackage{bbm}
\usepackage{cancel}

\DeclareGraphicsExtensions{.pdf,.gif,.jpg,.png,.eps}

\newcommand{\be}{\begin{equation}}
\newcommand{\beq}{\begin{equation}}
\newcommand{\ee}{\end{equation}}
\newcommand{\bea}{\begin{eqnarray}}
\newcommand{\eea}{\end{eqnarray}}
\newcommand{\ba}{\begin{array}}
\newcommand{\ea}{\end{array}}

\newcounter{saveeqn}

\def\d{\delta}
\def\D{\Delta}

\def\ve{\varepsilon}

\def\Tr{{\rm Tr}}

\def\1op{\hat{\mathbbm{1}}}
\def\1{\mathbbm{1}}

\begin{document}  

\title{Density Functional Theory of the Seebeck coefficient in the Coulomb 
blockade regime}

\author{Kaike Yang}
\affiliation{Nano-Bio Spectroscopy Group and European Theoretical Spectroscopy Facility (ETSF), Departamento de F\'isica de Materiales, Universidad del Pa\'is Vasco UPV/EHU, Av. de Tolosa 72, 
E-20018 San Sebasti\'an, Spain}

\author{Enrico Perfetto}
\affiliation{Dipartimento di Fisica and European Theoretical Spectroscopy Facility (ETSF), Universit\`{a} di Roma Tor Vergata,
Via della Ricerca Scientifica 1, 00133 Rome, Italy}
\affiliation{INFN, Laboratori Nazionali di Frascati, Via E. Fermi 40, 00044 Frascati,
Italy}

\author{Stefan Kurth}
\affiliation{Nano-Bio Spectroscopy Group and European Theoretical Spectroscopy Facility (ETSF), Departamento de F\'isica de Materiales, Universidad del Pa\'is Vasco UPV/EHU, Av. de Tolosa 72, 
E-20018 San Sebasti\'an, Spain}
\affiliation{IKERBASQUE, Basque Foundation for Science, E-48013, Bilbao, Spain}

\author{Gianluca Stefanucci}
\affiliation{Dipartimento di Fisica and European Theoretical Spectroscopy Facility (ETSF), Universit\`{a} di Roma Tor Vergata,
Via della Ricerca Scientifica 1, 00133 Rome, Italy}
\affiliation{INFN, Laboratori Nazionali di Frascati, Via E. Fermi 40, 00044 Frascati,
Italy}

\author{Roberto D'Agosta}
\affiliation{Nano-Bio Spectroscopy Group and European Theoretical Spectroscopy Facility (ETSF), Departamento de F\'isica de Materiales, Universidad del Pa\'is Vasco UPV/EHU, Av. de Tolosa 72, 
E-20018 San Sebasti\'an, Spain}
\affiliation{IKERBASQUE, Basque Foundation for Science, E-48013, Bilbao, Spain}

\begin{abstract}
The Seebeck coefficient plays a fundamental role in identifying the
efficiency of a thermoelectric device. Its theoretical evaluation
for atomistic models is
routinely based on Density Functional Theory calculations combined
with the Landauer-B\"uttiker approach to quantum transport. This 
combination,
however, suffers from serious drawbacks for devices in the Coulomb 
blockade regime. 
We show how to cure the theory through a simple correction in terms 
of the {\em temperature derivative} of   
the exchange correlation potential. Our
results compare well with both rate equations and experimental 
findings on
carbon nanotubes. 
\end{abstract}

\pacs{71.15.Mb,73.50.Lw,65.80.-g}

\date{\today}
\maketitle

The quest for increasingly energy-efficient technologies has recently led to 
significant scientific and technological interest in thermoelectricity \cite{DiSalvo1999,Vining2008,Vining2009}. 
Indeed, thermoelectric devices convert waste heat to electric power:
the basic working principle at their heart is the Seebeck
effect~\cite{Goldsmid2010}. The corresponding Seebeck coefficient is an 
important ingredient in the thermoelectric figure-of-merit (an
efficiency measure of a thermoelectric device). At present, the method of 
choice for an atomistic modelling of the Seebeck and other transport 
coefficients is density functional theory (DFT) combined with the 
Landauer-B\"uttiker formalism (LB-DFT)~\cite{DiVentra2008,DAgosta2013}.
However, an incautious use of LB-DFT as guide to
material and system selection may point in the wrong direction. In fact, LB-DFT 
is unable to capture the ubiquitous Coulomb blockade (CB) phenomenon 
of quantum devices weakly coupled to leads, thereby overestimating 
the conductance and, as we shall see, underestimating the Seebeck 
coefficient. In Ref.~\cite{KurthStefanucci:13,Liu:15,StefanucciKurth:15} it 
was shown that the erroneous high conductance predicted by LB-DFT stems
from neglecting 
exchange-correlation (xc) corrections to the bias
\cite{StefanucciAlmbladh:04,StefanucciAlmbladh:04-2,szvv.2005,kbe.2006,Vignale2009}. 
According to a recently proposed DFT 
framework for thermal transport (and thus for the calculation of the Seebeck 
coefficient)~\cite{EichVentraVignale:14,EichPrincipiVentraVignale:14}
xc corrections to the temperature gradient are also expected to occur.

In this Letter we propose an alternative DFT approach to the
Seebeck coefficient well suited for quantum devices in the CB regime.
Following a recent idea on the construction of xc corrections 
to the conductance~\cite{KurthStefanucci:13}, we find a very simple xc
correction to the LB-DFT Seebeck coefficient in terms of static DFT 
quantities. 
To illustrate the theory we first consider the Anderson impurity 
model (AIM), a paradigm for the CB effect~\cite{Costi2010}, and subsequently extend the 
analysis to multiple level systems. The proposed equations
are validated by benchmarking the results against those of the rate 
equations~\cite{Beenakker:91,BeenakkerStaring:92,Zianni2008} (RE),  
demonstrating the crucial role of the xc correction.
Finally, we apply the theory to single-wall carbon nanotubes
and find good qualitative 
agreement with experiment.

The Seebeck coefficient $S$ is defined as the ratio $S=(\Delta V/\Delta
T)_{I=0}$, where $\Delta V$ is the voltage that must be applied to cancel the
current $I$ generated by a small temperature difference $\Delta T$ between the
left and right leads. For an AIM symmetrically coupled to
featureless leads the Seebeck coefficient takes the form~\cite{Dong:02} (atomic units are used throughout)
\begin{equation}
S = -\frac{1}{T} \frac{\int \omega f'(\omega) A(\omega)}{\int 
f'(\omega) \, A(\omega)}, \quad \quad \int \equiv \int_{-\infty}^\infty \frac{d{\omega}}{2\pi} \;, 
\label{seebeck_siam}
\end{equation}
with $A(\omega)$ the interacting spectral function, $f(\omega) =1/(1 + e^{\beta
(\omega-\mu)})$ the Fermi function at temperature $T=1/\beta$ and chemical potential
$\mu$, and $f' \equiv d f/d \omega$. We now show how to rewrite $S$ in terms
of quantities which are {\em all} accessible by DFT. The starting point is the
equation $N=2\int f(\omega) A(\omega)$ for the electron occupation at the impurity.
Taking into account that, away from the particle-hole symmetric point, $N$ is an invertible function of $T$ we have $d A/d
T = (d A/d N)(d N/d T)$. Therefore, after some algebra we find
\begin{equation}
\frac{d N}{dT} = - \frac{2}{T} \frac{\int \omega f'(\omega)  
A(\omega)}{1+R}\;,
\label{dn_dtemp}
\end{equation}
where we have defined $R = -2 \int f(\omega) d A(\omega)/d N$. At the particle-hole
symmetric point, we have $dN/dT=0$ and $\int \omega f'(\omega)A(\omega)=0$
due to the fact that $A(\omega)$ is an even function of $\omega$, and thus Eq.~(\ref{dn_dtemp}) remains valid. With similar
steps we can also express the compressibility as
\begin{equation}
\frac{d N}{d\mu} = 
- \frac{2 \int f'(\omega) A(\omega)}{1+R}\;.
\label{dn_dmu}
\end{equation}
The combination of Eqs.~(\ref{dn_dtemp}) and (\ref{dn_dmu}) then gives
\begin{equation}
S = - \frac{d N/dT}{d N/d\mu}.
\label{seebeck_densderiv}
\end{equation}
This rewriting of the Seebeck coefficient is extremely interesting from a DFT
perspective since it involves exclusively the occupation $N$ of the {\em
equilibrium} system. We then calculate the derivatives in
Eq.~(\ref{seebeck_densderiv}) using the Kohn-Sham (KS) expression $N=2\int
f(\omega) A_{s}(\omega)$ where $A_{s}$ is the KS spectral function. 
For a gated impurity with
energy $v$ we have $A_{s}(\omega)=\ell(\omega-v-v_{\rm Hxc})$, 
with $\ell(\omega)=\gamma/(\omega^{2}+\gamma^{2}/4)$ a Lorentzian of
width determined by the
dot-lead tunneling rate $\gamma/2$. Thus, the dependence of $A_{s}$ on $N$ and
$T$ occurs only through the Hartree-xc (Hxc) potential $v_{\rm Hxc}$. By calculating the
required density derivatives and taking into account that $\frac{d v_{\rm
Hxc}}{d T}=\left(\frac{\partial v_{\rm Hxc}}{\partial 
N}\right)_{T}\frac{d N}{d
T}+\left(\frac{\partial v_{\rm Hxc}}{\partial T}\right)_{N}$, we
obtain the {\em exact} relation
\begin{equation}
S = S_s + \left(\frac{\partial v_{\rm Hxc}}{\partial T}\right)_{N} \; .
\label{seebeck_corr}
\end{equation}
Here $S_s$ is the KS Seebeck coefficient obtained from Eq.~(\ref{seebeck_siam})
by replacing $A(\omega)$ with $A_s(\omega)$, and it is exactly the coefficient
predicted by the LB-DFT approach.

Equation~(\ref{seebeck_corr}), the central result of this Letter, provides a rigorous route to cure
LB-DFT through
the inclusion of the xc correction $\partial v_{\rm Hxc}/\partial T$ while still
remaining in a pure DFT framework. As we shall see, Eq.~(\ref{seebeck_corr})
also suggests how to correct $S_{s}$ in larger systems.
\begin{figure}[t]
\includegraphics[width=8cm]{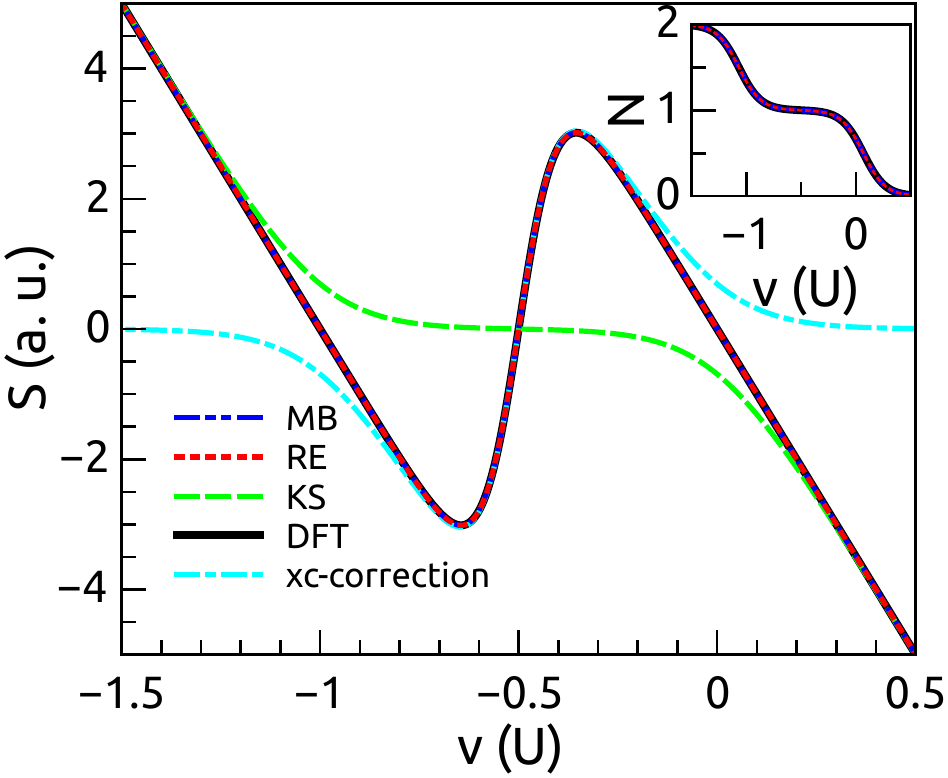}
\caption{(Color online) Seebeck coefficient $S$ and density $N$ (inset) 
versus gate $v$ for our corrected DFT (black), MB 
(blue) and RE (red). The $S_{s}$ (KS, green) and the xc correction
$\partial v_{\rm Hxc}/\partial T$ (cyan) are also displayed. The 
parameters are $T=0.1$ and  $\gamma=0.01$ (energies in units of $U$).}
\label{Siamfig}
\end{figure}

At temperatures $T\gg \gamma$, but still $T\ll U$ where $U$ is the on-site repulsion,
the CB phenomenon leaves clear fingerprints 
on the Seebeck coefficient. Nevertheless, these 
 are only partially captured by 
$S_{s}$, even when the exact $v_{\rm Hxc}$ is used (see below). 
The Anderson model is particularly instructive 
since it allows to disentangle the coordinated actions of the CB effect on 
$S_{s}$ and of the xc correction in reproducing the interacting $S$.

In the following we assume that $\gamma$ is the smallest energy scale
and we approximate $v_{\rm Hxc}$ by the exact  Hxc potential of the isolated ($\gamma=0$) impurity 
\cite{StefanucciKurth:11,PerfettoStefanucci:12},
\begin{equation}
v_{\rm Hxc}[N] \approx v_{\rm Hxc}^{\rm imp}[N] = \frac{U}{2} + g_{U}(N-1) \;,
\label{hxcpot_ss}
\end{equation}
where $g_{U}(x) = \frac{U}{2} + \frac{1}{\beta} \ln \left( 
\frac{x + \sqrt{x^2 + \exp(-\beta U)(1-x^2)}}{1+x} \right)$.
At low temperatures, the Hxc potential exhibits a sharp (but continuous) step
of size $U$ at occupation $N=1$ \cite{StefanucciKurth:11,Troster2012,Bergfield2012}.
With an analytic expression 
for $v_{\rm Hxc}$  we can evaluate both terms on the r.h.s. of 
Eq.~(\ref{seebeck_corr}). In Fig.~\ref{Siamfig} 
we show $S$ calculated from our DFT equation  (black) versus
the gate $v$. To demonstrate the accuracy of the result we also show the Seebeck 
coefficient calculated from Eq.~(\ref{seebeck_siam}) using the 
Many-Body (MB) spectral function
$A(\omega) =\frac{N}{2}\ell(\omega-v-U) + \left( 1- \frac{N}{2}\right)
\ell(\omega-v)$~\cite{HaugJauho:08} (blue) 
as well as the one calculated using the RE approach of 
Ref.~\cite{BeenakkerStaring:92} (red), exact in the 
limit $\gamma \to 0$.
All three approaches give the same Seebeck coefficient 
and densities (see inset). Let us now discuss how 
the two terms in Eq.~(\ref{seebeck_corr}) contribute.
The KS Seebeck $S_{s}$ (green) accounts for the correct 
linear behavior (with slope proportional to $T^{-1}$) at large 
values of $|v|$. In fact, for $\gamma\to 0$ the KS spectral function 
becomes $A_s(\omega) = 2 \pi 
\delta(\omega -v -v_{\rm Hxc})$ and consequently $S_s = -(v+v_{\rm Hxc})/T$. 
The linear behavior at large $|v|$
is not surprising since the noninteracting Seebeck coefficient 
behaves in the same way. Noteworthy is instead the 
plateau of $S_{s}$ for $v\in (-U,0)$. This is a direct 
consequence of the step in $v_{\rm Hxc}$ which pins the KS 
level to the chemical potential thereby blocking electrons 
with energy below $v+U$ from entering the impurity site (see inset).
The CB-induced plateau in $S_{s}$ opens a gap in the 
noninteracting straight line $-v/T$, shifting it leftward by $U$ for 
$v<-U$ and generating the correct behavior at large negative values 
of $v$. However, 
$S_{s}$ misses entirely the oscillation of $S$ for $N\approx 1$, 
thus severely underestimating the true Seebeck coefficient.
Remarkably, this deficiency is exactly cured by 
the xc correction $\partial v_{\rm Hxc}/\partial T$ (cyan).
The  temperature variation of $v_{\rm Hxc}$
is the key ingredient for the nonvanishing Seebeck 
coefficient in the CB regime~\cite{BeenakkerStaring:92,StaringMolenkamp:93,Dzurak:93,Dzurak1997}.

We now extend the DFT approach to junctions with more than one level.
For $T\gg\gamma$ the Seebeck coefficient exhibits a sawtooth behavior as 
a function of $v$, with ``jumps'' occurring when the number $N$ of 
electrons crosses an integer. 
Furthermore, if the level spacing $\D\ve$ is much larger than $T$,  
a superimposed fine structure of wiggles spaced by $\D\ve$ 
emerges~\cite{BeenakkerStaring:92}. The wiggles originate from 
excitations that bring the system with $(N-1)$ particles in the 
ground state to some excited state with $N$ particles.

The physics of the Seebeck coefficient in a multiple level junctions 
is well captured by 
the Costant Interaction Model (CIM). 
The CIM Hamiltonian reads 
$\hat{H}=\sum_{i\sigma}\ve_{i}\hat{n}_{i\sigma}+\frac{1}{2}\sum_{i\sigma\neq 
j\sigma'}U_{ij}\hat{n}_{i\sigma}\hat{n}_{j\sigma'}$, 
where $\hat{n}_{i\sigma}$ is the occupation operator of the $i$-th level 
with spin $\sigma$. The indices $i,j$ run over $M$ levels and for 
$M=1$ we are back to the Anderson model.
For simplicity we assume that each level is equally 
coupled to the left and right leads with tunneling rate $\gamma/2$. In 
this case the derivation of Eq.~(\ref{seebeck_densderiv}) can be repeated
step by step by replacing the spectral function $A$ with its 
trace $\Tr[A]$. Consequently, we can again express $S$ in a pure DFT 
framework by calculating the derivatives of the total number of 
electrons from the KS expression $N=2\int 
f(\omega)\Tr[A_{s}(\omega)]$. The KS spectral function 
$[A_{s}]_{ij}=\d_{ij}A_{s,i}$ is diagonal in the level basis and 
reads $A_{s,i}(\omega)=\ell(\omega-\varepsilon_i-v_{{\rm Hxc},i})$, where the Hxc 
potential of level $i$ depends on the 
occupations $\{n\}$ of all the levels. It is straightforward to 
show that
\begin{equation}
S=S_{s}+\sum_{j}
\frac{\int f'(\omega)A_{s,j}(\omega)}{\int 
f'(\omega)\Tr[A_{s}(\omega)]}
\left(\frac{\partial v_{{\rm Hxc},j}}{\partial T}\right)_N.
\label{mult_level_seebeck}
\end{equation}

In Ref.~\onlinecite{StefanucciKurth:13} we proved that at zero 
temperature the Hxc potential of the isolated ($\gamma=0$)
CIM Hamiltonian is uniform and depends only on $N$, 
i.e., $v_{{\rm Hxc},i}[\{n\}]=v_{\rm Hxc}[N]$. The zero-th order 
approximation at finite temperatures and weak coupling 
to the leads 
therefore consists in neglecting the nonuniformity and the local
dependence on the $\{n\}$. In this approximation 
Eq.~(\ref{mult_level_seebeck}) reduces to Eq.~(\ref{seebeck_corr}).
The interacting Seebeck coefficient then
follows once we specify the functional form of $v_{\rm Hxc}$.
Following Ref.~\onlinecite{KurthStefanucci:13} 
we construct $v_{\rm Hxc}[N]$ as the sum of single-impurity Hxc potentials 
according to
\begin{equation}
v_{\rm Hxc}[N] = \sum_{K=1}^{2 M -1} \left[\frac{U_{K}}{2}+
g_{U_{K}}^{\rm ext}(N-K)\right],
\label{vxc_multi_sum}
\end{equation}
where $U_{K}$ is the charging energy needed for adding one electron to 
the system with $K$ electrons, and the extended $g^{\rm ext}_{U}$ function 
is defined according to
\begin{equation}
g_{U}^{\rm ext}(N-1) = \left\{ 
\begin{array}{cl}
-U/2 & \mbox{ $N<0$} \\
g_{U}(N-1) & \mbox{ $0\leq N \leq 2$}~~, \\
U/2 & \mbox{ $N>2$} 
\end{array}
\right.
\end{equation}
with $g_{U}$ given below Eq.~(\ref{hxcpot_ss}). The Hxc potential in 
Eq.~(\ref{vxc_multi_sum}) has a staircase behavior with steps of 
width $U_{K}$ between two consecutive integers. 
\begin{figure}[tbp]
    \includegraphics[width=8cm]{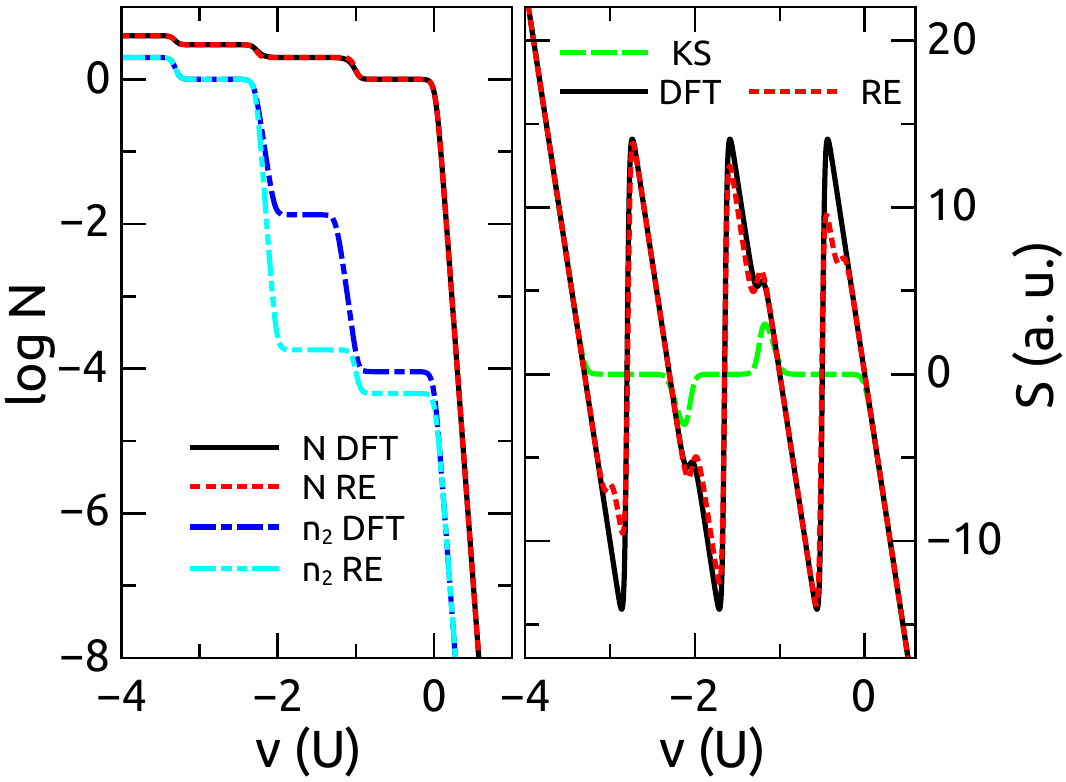}
    \caption{(Color online) Density (left) and Seebeck coefficient (right)
of CIM with two spin-degenerate levels computed from RE and DFT using the
approximate functional of Eq.~(\protect\ref{vxc_multi_sum}). The KS Seebeck
coefficient is also shown.}
\label{homo_lumo}
\end{figure}

\begin{figure}[tbp]
\includegraphics[width=8cm]{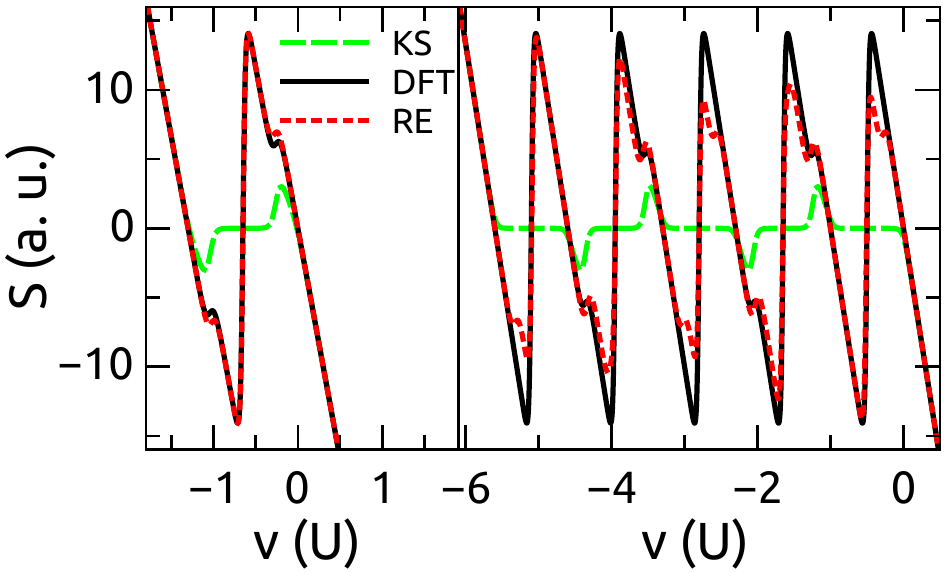}
\caption{(Color online) Seeebeck coefficient for the Anderson model with 
non-degenerate single-particle levels (left) and for three 
spin-degenerate levels (right). The parameters are 
$\varepsilon_{\uparrow}^0=0$, $\varepsilon_{\downarrow}^0=0.3$ (left 
panel) and
$\varepsilon_{1}^0=0$, $\varepsilon_{2}^0=0.3$, 
$\varepsilon_{3}^0=0.6$ (right panel). In 
both panels $T=0.03$ and $\gamma=0.001$ (all energies in units of $U$).}
\label{multiple}
\end{figure}

To assess the quality of our approximate Hxc potential we first 
consider a two-level CIM with $U_{ij}=U$.
In Fig.~\ref{homo_lumo} we display results 
at temperature $T=0.03$, coupling $\gamma=0.001$ and  
$\varepsilon_{i}=\varepsilon_{i}^{0}+v$ where $\varepsilon^{0}_{1}=0$ and $\varepsilon_{2}^{0}=0.3$
(energies in units of $U$).  The left panel shows the 
total occupation $N$ as well as the  
occupation $n_{2}=\sum_{\sigma}n_{2\sigma}$ of the highest level
calculated using both DFT and RE. Although a perfect 
agreement is found for $N$, exponentially small discrepancies are 
seen for the local occupation. In fact, the uniformity (i.e., level independence) of our zero-th order 
approximation $v_{\rm Hxc}$ of Eq.~(\ref{vxc_multi_sum}) 
neglects thermal excitations, which 
corresponds to mixing only ground states of different 
$N$. Accordingly, the DFT Seebeck coefficient is expected to exhibit 
only those wiggles associated with the addition of one  
electron in the lowest available level. This is confirmed by 
the right panel of Fig.~\ref{homo_lumo} where the wiggles associated to 
the addition energies of excited states are captured by 
the RE (red) but missed by DFT (black). For a perfect agreement between DFT and RE one should abandon
the zero-th order uniform approximation and consider a level-dependent Hxc 
potential which correctly reproduces the level occupations. 
To further support this analysis on the relation between the 
nonuniformity of $v_{\rm Hxc}$ and physical excitations
we show in Fig.~\ref{multiple} 
the Seebeck coefficient for the Anderson model with 
broken spin degeneracy (left) and for a three-level 
CIM (right). In the first case DFT agrees with RE since there exists 
only one addition energy, whereas in the second case DFT misses the 
wiggles of excited-state addition energies. We emphasize that the 
wiggles stem from $S_{s}$ (green line), and are not due to the xc 
correction. The latter is responsible for the large sawtooth 
oscillations and, as Figs.~\ref{homo_lumo} and~\ref{multiple} clearly 
show, it is the dominant contribution to $S$. 

Recently experimental measurements of the Seebeck coefficient and the electrical conductance in the 
CB regime have been reported for an individual single-wall carbon 
nanotube~\cite{Small2003} as well as for quantum dots~\cite{Dzurak1997,Dzurak:93}. 
For the transport properties of nanotubes, we can extract from the experimental results both single-particle energies and 
charging energies which are then used to 
calculate both $G$ and $S$ with our DFT scheme. In contrast to the model 
calculations described previously, here the charging energies $U_K$ depend on the charging state $K$.

\begin{figure}[t!]
\includegraphics[width=8cm]{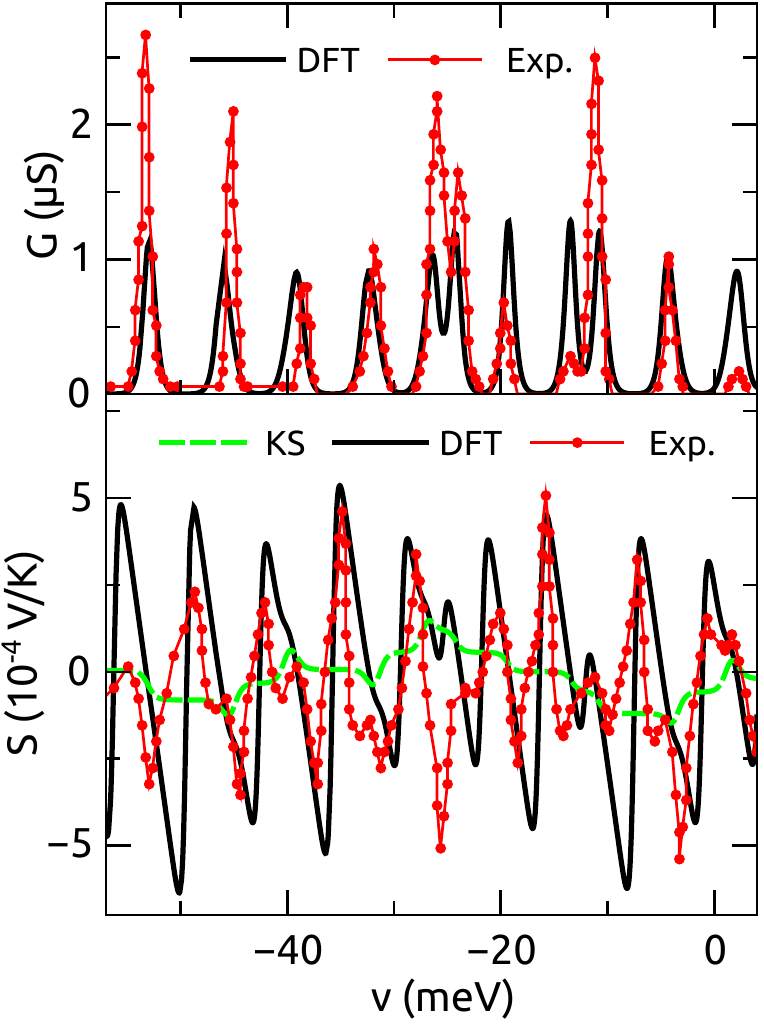}
\caption{(Color online) Conductance (upper panel) and Seebeck coefficient 
(lower panel) of a single-wall carbon nanotube from DFT (black) and 
experiment [red, data from Ref.~\onlinecite{Small2003}]. 
Also shown is the KS Seebeck coefficient (dashed green). 
The single particle and charging energies are given in 
Table~\ref{spenergies_charging}. The other 
parameters are $T=4.5$ K and $\gamma=0.02$ meV.}
\label{CNT}
\end{figure}

\begin{table}[ht]
\label{Ec}
\caption{Single-particle energies $\varepsilon^0$ and charging energies $U_K$ 
(in meV), for modelling the calculation of Fig.~\ref{CNT}.}
\begin{ruledtabular}
\begin{tabular}{l|cccccc}
$\varepsilon^0$ & -6.0 & -3.75 & -3.75 & -3.75 & -1.5 & 0.75 \\ \hline
$K$, $U_K$ 
& 
1, 3.75
& 
\begin{tabular}{c}
2, 5.0 \\
3, 6.25
\end{tabular} & 
\begin{tabular}{c}
4, 2.25 \\
5, 5.75
\end{tabular} & 
\begin{tabular}{c}
6, 4.5 \\
7, 2.0
\end{tabular} & 
\begin{tabular}{c}
8, 4.5 \\
9, 6.5
\end{tabular} 
& 
\begin{tabular}{cc}
10, 5.25 \\
11, 6.75
\end{tabular}
\end{tabular}
\end{ruledtabular}
\label{spenergies_charging}
\end{table}
In Fig.~\ref{CNT} we present both the conductance $G$ (upper panel), 
and the Seebeck coefficient (lower panel) 
$S$ as a function of gate voltage $v$ calculated with the parameters 
listed in Table \ref{spenergies_charging} and for temperature $T=4.5$ 
K and $\gamma=0.02$ meV. For comparison we also report the Seebeck 
coefficient as calculated from the LB-DFT formalism with the same 
parameters. LB-DFT fails in reproducing the characteristic sawtooth 
behaviour of the experimental results. Instead, the Seebeck 
coefficient calculated with our DFT scheme clearly shows the peak       
and valley structures observed in experiment, confirming again the crucial 
role of the xc correction.
Also, all the fine 
structure wiggles (kinks in some cases) are correctly captured.

In conclusion, we have proposed a DFT scheme for the calculation of the Seebeck 
coefficient which corrects the deficiencies of the canonical 
Landauer-B\"uttiker approach in the Coulomb blockade regime. 
We found that two ingredients in the Hxc potential are essential:
(i) the step feature at integer electron 
number opens a gap in the linear dependence on gate voltage and  
(ii) the temperature derivative generates the sawtooth behaviour 
in this gap region.
Remarkably, the xc correction represents the 
dominant contribution to $S$ just as the xc correction to the 
conductance dominates in the Coulomb blockade 
regime~\cite{KurthStefanucci:13}. We have compared our theory with 
both rate equations
and experimental results on a carbon
nanotube, and found good quantitative agreement in all cases. 
The present approach is valid in the linear response regime, where the 
applied thermal gradient is small. Going beyond the linear response would pave the
way for a deeper understanding of the thermoelectric effect 
and allow to study materials for extreme applications. The recently 
proposed
DFT framework for thermal transport by Eich et al.~\cite{EichVentraVignale:14}
appears a promising starting point for this purpose. 

\begin{acknowledgments}
We would like to acknowledge useful discussions with F. Eich at the early stage of this project.
K. Y., S. K., and R. D'A. acknowledge financial support from DYN-XC-TRANS 
(Grant No.~FIS2013-43130-P) and NANOTherm (CSD2010-00044) of the Ministerio de
Economia y Competitividad (MINECO), and the Grupo Consolidado UPV/EHU del 
Gobierno Vasco (IT578-13). E. P. and G. S. acknowledge funding by MIUR FIRB Grant No. RBFR12SW0 and EC funding through the RISE CoExAN (GA644076).
\end{acknowledgments}

\end{document}